\documentclass[prb,twocolumn,floatfix,superscriptaddress]{revtex4}

\usepackage{lscape}
\usepackage{rotating}
\usepackage{epstopdf}
\usepackage{graphicx}
\usepackage{natbib}
\usepackage{epstopdf}
\usepackage{amsmath}
\usepackage{amssymb}
\usepackage{mathbbol}
\usepackage{xcolor}
\usepackage{lipsum}
\usepackage{amssymb}
\usepackage{mwe}
\usepackage{lipsum} 
\usepackage{siunitx}
\usepackage[normalem]{ulem}

\usepackage{empheq}
\usepackage{afterpage}
\usepackage[colorlinks=true,citecolor=blue,linkcolor=blue,urlcolor=blue]{hyperref}
\usepackage{cleveref}

\newcommand{\beq}{\begin{equation}}
	\newcommand{\eeq}{\end{equation} \smallskip}
\newcommand{\beqy}{\begin{eqnarray}}
	\newcommand{\eeqy}{\end{eqnarray} \smallskip}

\newcommand{\bit}{\begin{itemize}}
	\newcommand{\eit}{\end{itemize}}
\newcommand{\bmat}{\begin{pmatrix}}
	\newcommand{\emat}{\end{pmatrix}}

\begin{document}
	
\title{Coherent transfer of topological domain walls}

\author{P. Comaron}
\affiliation{Institute of Physics, Polish Academy of Sciences, Al. Lotnikow 32/46, 02-668 Warsaw, Poland}

\author{V. Shahnazaryan}
\affiliation{Institute of Physics, Polish Academy of Sciences, Al. Lotnikow 32/46, 02-668 Warsaw, Poland}
\affiliation{ITMO University, St. Petersburg 197101, Russia}

\author{M. Matuszewski}
\affiliation{Institute of Physics, Polish Academy of Sciences, Al. Lotnikow 32/46, 02-668 Warsaw, Poland}

\date{\today}

\begin{abstract}
We demonstrate the controlled coherent transfer of topological domain walls in a one-dimensional non-Hermitian chain of interacting Bose-Einstein condensates. The topological protection stems from a spatially patterned pump in an open-dissipative system. As a test bed setup of the proposed phenomenon, we consider a chain of coupled micropillars with embedded quantum wells, possessing exciton-polariton resonances. The transfer of a domain wall is driven by spatially localised, adiabatic pump modulation in the vicinity of the domain wall. The stochastic calculations prove the coherent nature of the domain wall transfer. For appropriate system parameters the coherence degree is preserved after multiple transitions, paving the way towards long-range transfer of a coherent quantum state.
\end{abstract}

\maketitle


\section{{Introduction}}

Topological insulators (TI) are a class of materials that possess an energy bandgap and topologically protected low energy states~\cite{HasanReview,Chiu2016}. Topological protection in these systems stems from the symmetry of the bulk, which is quantified by means of topological invariants. Bulk-boundary correspondence results in the protection of edge states, which hold promise for applications in dissipation-less communications and quantum computing.

While in standard TIs non-trivial topology results from the properties of a Hermitian Hamiltonian, recently a class of non-Hermitian topological systems attracted great interest~\cite{Ueda_TopologicalPhasesNonHermitian}. These are of particular relevance to photonics, where open-dissipative effects are prevalent. The latter, on one side, makes photonic systems an optimal platform for probing phenomena emerging specifically in the non-Hermitian domain. In this context phenomena such as lasing of topological states have been demonstrated~\cite{Bahari2017,Bandres2018}. On the other side, recently it was proposed to use the non-Hermiticity of photonics as an efficient tool for controlling the topological properties of the system. This can be reached, particularly, through asymmetric coupling coefficients~\cite{Midya_NonHermitianPhotonics} or spatial modulation of gain-loss ratio in each site~\cite{Takata2018,Zhao2019,Comaron2019}.
Arguably, the greatest fundamental interest lies in the investigation of topological states that result solely from the non-Hermiticity of the system, since these have no counterparts in the Hermitian case~\cite{Leykam2017,Yao2018,YaoBand2018,Kunst2018,Shen2018, MATORRES18, Yokozimo2019, Ueda_TopologicalUnification}.

In this paper, we consider the question whether non-Hermiticity can be used for precise control of topological states. Similarly to electrons in crystalline media, electromagnetic waves in periodically patterned photonic structures form energy bands, which can lead to appearance photonic edge states~\cite{Hafezi2011,Hafezi2013,Zeuner2015,Zhan2017,Xiao2017,Weimann2017, Bahari2017, Parto2018, Bandres2018, Zhao2018,Zhou2018,Olekhno2020}. The presence of topological protection suppresses the backscattering on disorder, thus generating an energy-efficient propagation channel. However, in this simple setting there is no control over the direction or velocity of the wave packet. On the other hand, photonic implementations of non-Hermitian Hamiltonians relying on spatial modulation of external pumping allow to efficiently tune certain terms of the Hamiltonian. This opens the way to the control of topological states, which can be crucial for future applications, such as scattering-free optical interconnects, quantum computation, or Majorana state braiding~\cite{DasSarma_MajoranaZeroModes}.

The distinctive peculiarities of photonic TIs can be further extended in the regime of strong-light matter coupling in a microcavity with embedded quantum wells \cite{CarusottoReview}. The emerging hybrid quasiparticles, called exciton-polaritons, are interacting via their excitonic component, allowing thus to achieve a strong nonlinear response in comparison to other photonic systems.
Typically, etching of cavity is used to fabricate an array of coupled micropillars, mimicking the structure of a tight-binding Hamiltonian. In the majority of existing theoretical proposals \cite{Bardyn2015,Bleu2016,Banerjee2018,Janot_TopologicalHallCavity,Malpuech_TopologicalGapSoliton,Savenko_TIMagneticDots,Malpuech_ZTopologicalInsulator,Liew_FloquetTopologicalPolaritons,Shelykh_TopologicalMetamaterialsRings,Kartashov_2DTopologicalPolaritonLaser,Kartashov_BistableTI,Downing_TopologicalPhasesCavityWaveguide,Liew_ChiralBogoliubons,Liew_SpontaneousChiralEdgeStates,Sigurdsson2019} and experimental realizations~\cite{stJean2017,Bloch_TopologicalInvariantsQuasicrystals,Amo_OrbitalEdgeStates,klembt2018,Whittaker2018} topological order emerges from Hermitian band engineering, whereas open-dissipative nature of the system serves only to create a non-equilibrium Bose-Einstein condensate of polaritons in each micropillar.

As we demonstrated recently, topological protection in a chain of coupled polariton micropillars can be achieved solely via the spatial modulation of external pump \cite{Comaron2019} in a system with equal hopping coefficients. Topological characterization of the system revealed the existence of multiple phases, with different number of end states. Here, we show that at the boundary of such phases a non-decaying topological domain wall can be created. By means of adiabatic switching of the pump pattern, we induce a controllable transfer of the domain wall. Moreover, by calculating the first-order correlation function, we demonstrate that such transfer is of coherent nature. We determine the optimal conditions for the coherent transfer, such as time dependence of the spatial pump pattern and the switching time. Our results are confirmed by stochastic simulations within the truncated Wigner approximation, which include the effect of quantum  fluctuations.

\section{The model}

\subsection{Topological domain walls}

The structure we consider consists of a chain of unit cells, each including four sites. The on-site potential within each cell is spatially modulated. The bulk of the chain is analogous as in our previous work~\cite{Comaron2019}. Here, we consider domain walls which emerge at the boundary between two phases supporting a different number of edge states. The Hamiltonian of the system reads
\begin{align}
    \hat{H} =  e^{i \theta}  \sum_{n=1}^{n_b}  \left( 
	g_1 \hat{a}^\dagger_n \hat{a}_n  
	- g_2 \hat{b}^\dagger_n \hat{b}_n 
	- g_1 \hat{c}^\dagger_n \hat{c}_n 
	+ g_2 \hat{d}^\dagger_n \hat{d}_n 
	\right)   \nonumber \\
	+ e^{i \theta} \sum_{n=n_b+1}^{N}  \left( 
	g_3 \hat{a}^\dagger_n \hat{a}_n  
	- g_4 \hat{b}^\dagger_n \hat{b}_n 
	- g_3 \hat{c}^\dagger_n \hat{c}_n 
	+ g_4 \hat{d}^\dagger_n \hat{d}_n 
	\right)   \nonumber \\
    + \kappa \sum_{n=1}^N  \left( 
    \hat{b}^\dagger_n \hat{a}_n + \hat{c}^\dagger_n \hat{b}_n + \hat{d}^\dagger_n \hat{c}_n + \hat{a}^\dagger_{n+1} \hat{d}_n 
    + H.c. \right) ,
	\label{eq:Hamiltonian}
\end{align}
where $N$ is the total number of unit cells, and $n_b$ denotes the boundary unit cell between the two phases. Here $g_i e^{i\theta}$ denotes the on-site potential, which is generally a complex valued quantity. $\kappa$ denotes the nearest neighbour hopping rate between the sites, which is uniform throughout the chain. Such a Hamiltonian corresponds, in particular, to an exciton-polariton system of coupled micropillars, where the imaginary part of the on-site potential results from an external incoherent pump, and the real part stands for the Coulomb interaction between particles in a polariton condensate and in an incoherent reservoir \cite{Comaron2019}. The parameter $\theta$  describes the ratio  of  real  and  imaginary  components  of  the  on-site  potential, and thus is determined by the material composition and design of the system, including the exciton-polariton detuning.

\begin{figure}[t]
	\centering
	\includegraphics[width=\linewidth]{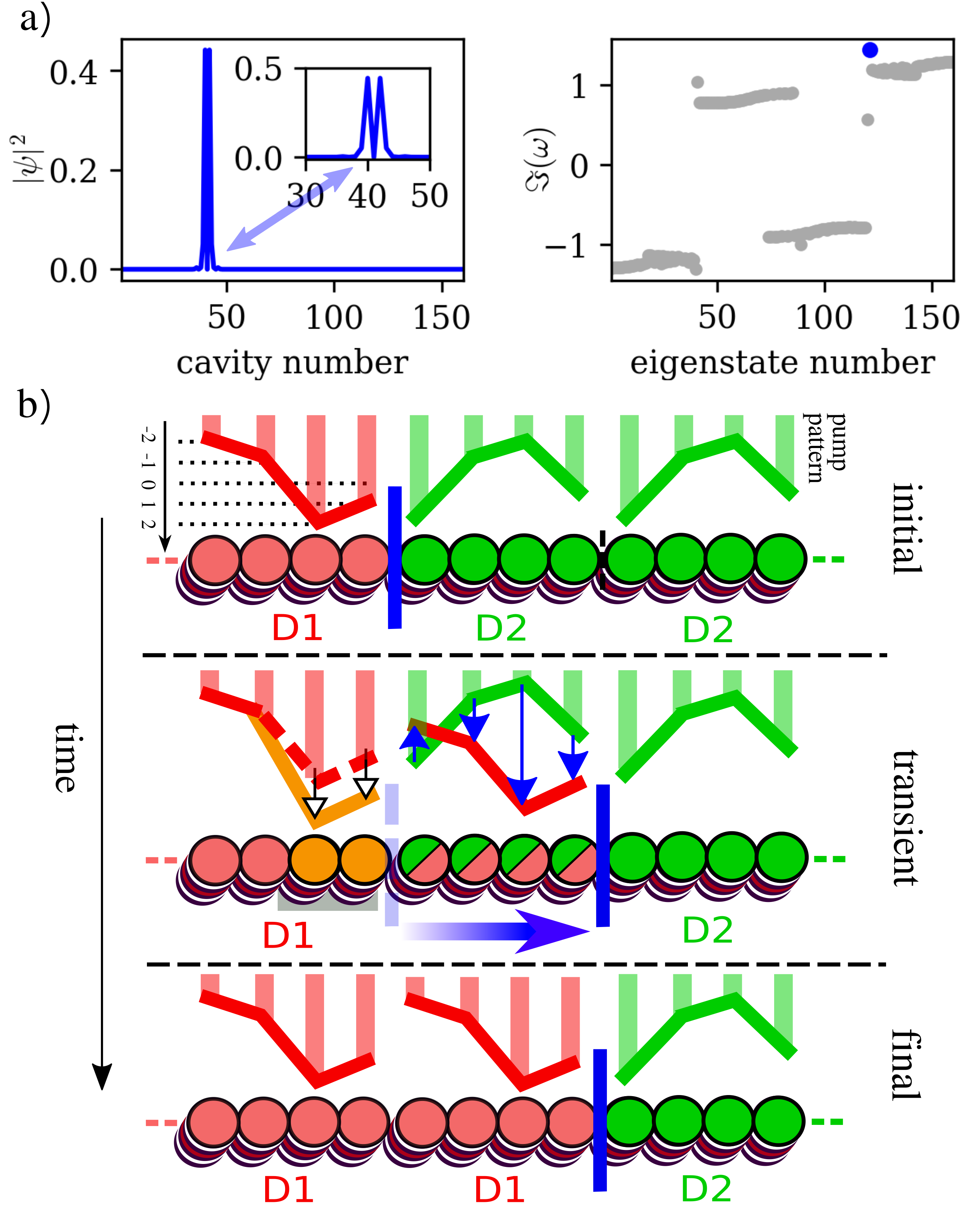}
	\caption{
	(a) The spatial density distribution of an eigenstate corresponding to a topological domain wall emerging on the boundary between two topological phases (left panel). The imaginary parts of eigenvalues of Hamiltonian \eqref{eq:Hamiltonian} in units of $\kappa$ (right panel). The blue dot corresponds to domain wall mode.
    (b) The sketch of boundary between the topological phases and its temporal evolution. In transient regime the pump rates in middle unit cell are gradually modified, resulting in shifting the boundary to one unit cell. In the transient regime a supporting potential is applied to the boundary pillars (marked by orange), which preserves the initial domain wall mode from depletion during the transfer of the state.
	}
	\label{fig:eigensystem}
\end{figure}

It was previously shown \cite{Comaron2019} that in the homogeneous case (i.e. $g_3=g_1$, $g_4=g_2$) the Hamiltonian \eqref{eq:Hamiltonian} can exhibit one or two pairs of edge states, odd number of edge states, or be topologically trivial. In order to perform dynamical study of domain wall behaviour in a dissipative system, one needs a single topological state to be the only non-decaying eigenstate. Hence, the imaginary part of domain wall eigenenergy needs to be the highest among all the eigenmodes.  Such a situation can be reached via judicious choice of parameters, with the domain wall appearing on the boundary of two topological phases, exhibiting one and two pairs of edge states, respectively. Particularly, 
here we choose the values of parameters $g_1=-2\kappa$, $g_3=2\kappa$, $g_2=g_4=\kappa$, $\theta=\pi/3$, the length of the chain $N=40$, and the boundary unit cell $n_b=10$. The shape of the domain wall state and the imaginary part of the energy spectrum are
shown in Fig.~\ref{fig:eigensystem}(a), left and right panels, respectively. Notably, the domain wall is ``M''-shaped, almost completely localized in 5 sites, and centred exactly around the boundary site between the two phases.
%


\subsection{  Protocol of the transition}

\begin{figure}[t]
	\centering
	\includegraphics[width=\linewidth]{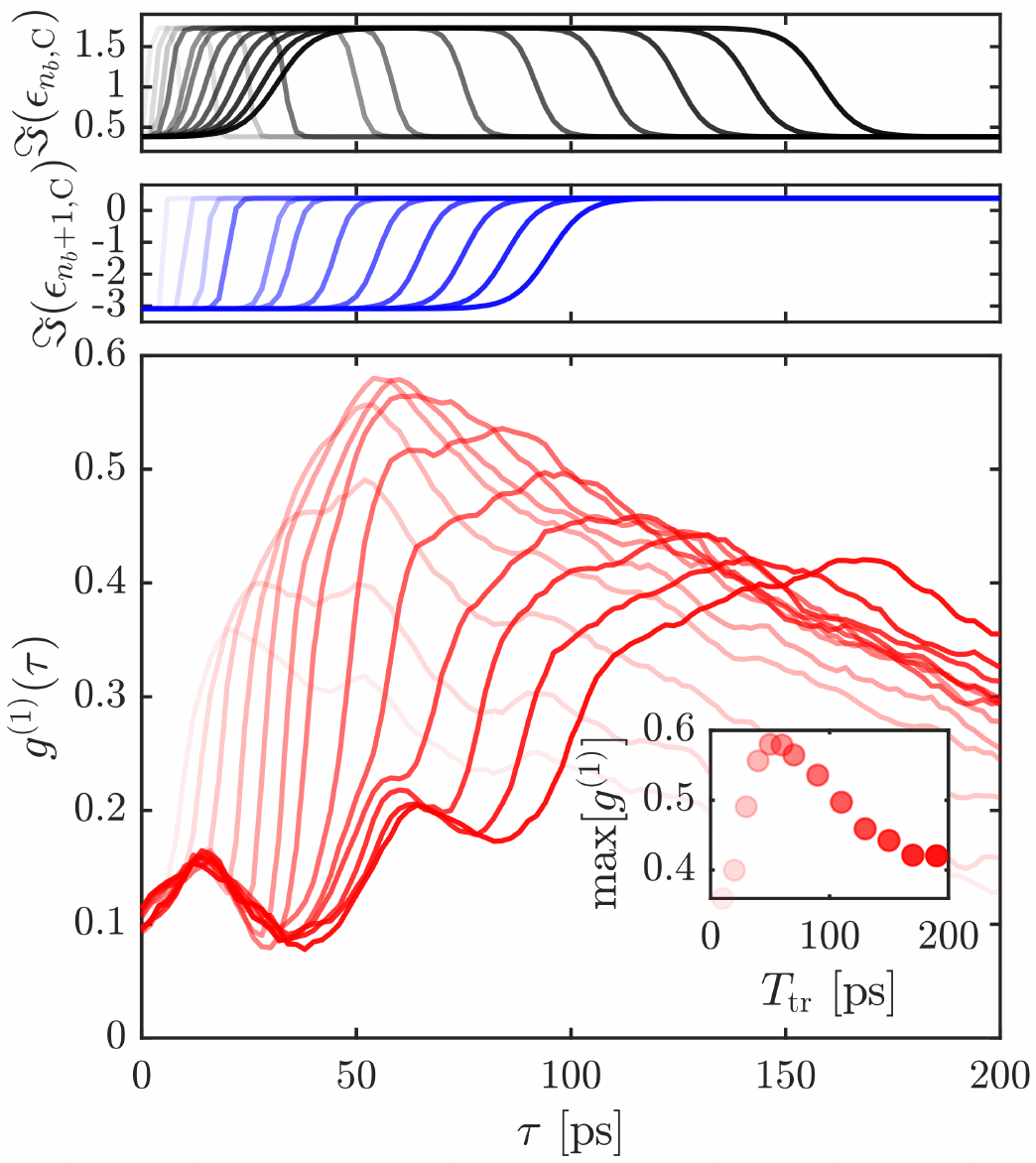}
	\caption{
	The dependence of domain wall coherence on switching speed in the case where domain wall is shifted by one unit cell. Top panel: the temporal dependence of supporting pump applied to the existing domain wall. Middle panel: temporal dependence of pump rates in the boundary unit cell. The pump values are in units of $\kappa$. Bottom panel: first order correlation function defined by Eq.~\eqref{eq:gonedef} for different transition times $T_\mathrm{tr}$. The inset illustrates the maxima of coherence at different transition times.
    The colour tones in plots correspond to values of $T_\mathrm{tr}$ as shown in inset.  
	The interplay between the adiabaticity rate and finite coherence time leads to coherence maxima appearing for intermediate transition time, which for the chosen parameters is $T_\mathrm{tr}=50$ ps.
	Hereafter  the scaling factor of noise is $\beta = 0.05$, $\kappa=0.1$ meV, and $\sigma=\kappa$. 
	}
	\label{fig:dynamics}
\end{figure}
\begin{figure}[t]
	\centering
	\includegraphics[width=\linewidth]{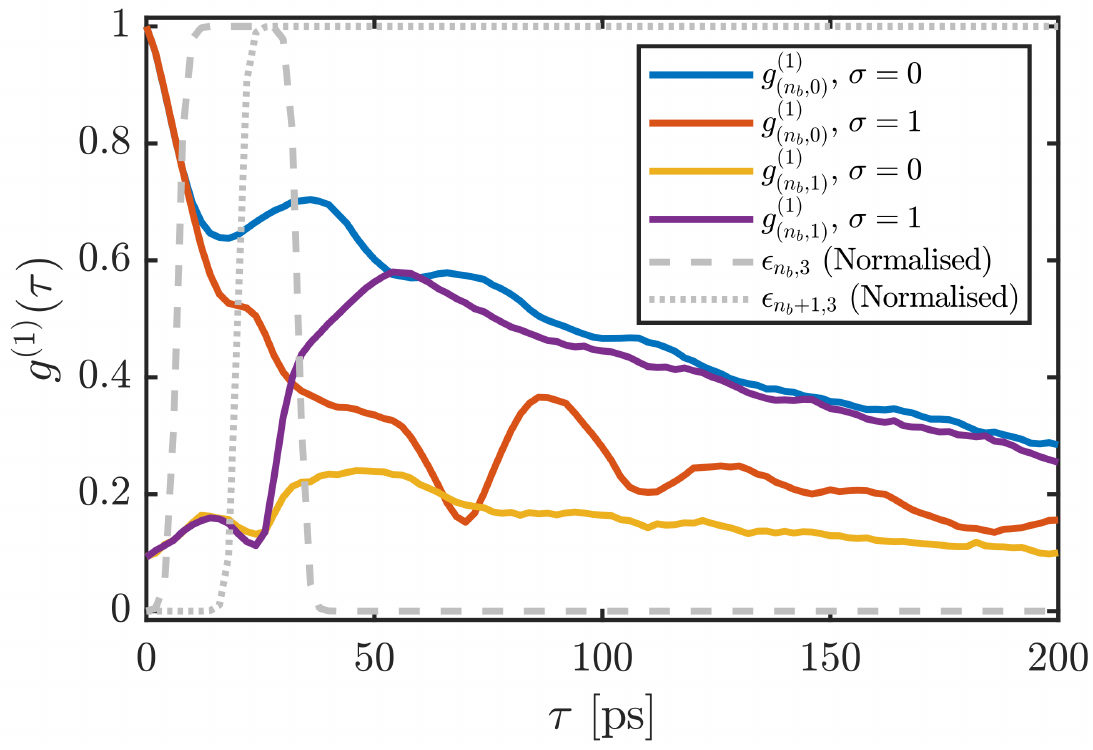}
	\caption{
	The influence of supporting pump on coherence. 
	The orange and purple curves show the evolution of correlation function in the absence and presence of supporting pump, respectively. The gray dashed and dotted lines correspond to temporal profiles of supporting pump and pump switching in the boundary unit cell.
	For comparison we show the evolution of local coherence in the scenario when no transition happens. The red curve corresponds to the presence of supporting pump, and the blue curve to its absence. The supporting pump in this case distorts the established steady state, and impose an additional noise. Hereafter $T_\mathrm{tr}=50$ ps.
	}
	\label{fig:protocols}
\end{figure}

We proceed with the study of coherent transfer of a domain wall. For the sake of purity we consider the case when the only non-decaying mode is represented by the domain wall. To achieve this, we uniformly reduce the on-site potentials of the system, so that only the domain wall mode has positive imaginary part of eigenvalue.
The protocol of transfer consists of several steps, which are illustrated as follows. 

For a better understanding, let us consider first the evolution in the mean field regime, as sketched in Fig.~\ref{fig:eigensystem}(b). 
We start with a random distribution, and eventually reach a nonequilibrium steady state configuration, if growth saturation is present. The latter naturally emerges in any system where pumping has a limited capacity.
This state is labelled as ``initial'' in Fig.~\ref{fig:eigensystem}~(b). 
Then, we gradually modify the pump pattern at the boundary of two phases, and thus shift the boundary by one unit cell.
In the beginning of the ``transient" stage the existing domain wall ceases to be a non-decaying mode, and rapidly becomes depleted. The new domain wall emerges at the new boundary between the phases, and becomes populated. 
Steady state of the configuration hosting the new shifted domain wall is eventually reached in the ``final" state.
Yet, such a combination of two independent processes cannot be considered as a true transfer of macroscopic population. To avoid this scenario, during the transition process we apply extra pump to the region of existing domain wall, which replenishes the decaying population. In this case we find that the domain wall is indeed shifted by one unit cell.

To verify this, we perform stochastic simulations, and calculate the spatially and temporally resolved degree of coherence between domain walls at the initial and final stages of the process. 
The evolution of condensate can be described by a stochastic discrete Gross-Pitaevskii equation \cite{Comaron2019}, in which sites are coupled to each other due to the presence of hopping. 
The corresponding set of equations
reads
\begin{align}
          i \hbar \mathrm{d}\psi_{n,i} = 
    \left[ - \kappa \sum_{\langle nn \rangle} \psi_{n',i'} +
    \epsilon_{n,i} (t) \psi_{n,i} \right. \nonumber \\
    \left. - \Gamma_{n,i}\left( 1 + i \tan \theta \right)|\psi_{n,i}|^2 \psi _{n,i} \right] \mathrm{d}t +\xi_{n,i}(t),
    \label{eq:dynamics}  
\end{align}
where $\langle nn \rangle$ runs over the nearest neighbours, $n\in [1,N]$ and $i=A,B,C,D$ is the sub-lattice label. Here $\psi_{n,i}(t)$ is the condensate amplitude in the corresponding site, $\epsilon_{n,i} (t) =g_i e^{i\theta} - i \gamma$, with the term $-i\gamma$ corresponding to the uniform reduction of the on-site potential (see Appendix~\ref{Sec:AppendixA} for details of the derivation).
The parameter $\Gamma_{n,i}$ describes the nonlinearity in each site. The last term denotes a Gaussian white noise with correlations $\langle \xi_{ni} (t) \xi^{*}_{n'i'} (t') \rangle =\delta_{nn'}\delta_{ii'} \delta_{tt'} \beta^2( g_i \sin \theta -\gamma +\gamma_c )/d  $ 
accounting for thermal and quantum fluctuations. It should be noted that the chosen noise amplitude corresponds to the particular case of exciton-polariton condensates, but in principle an analogous definition can be given for other related systems.
Here the parameter $\beta$ is the dimensionless scaling factor describing the scaling of noise amplitude, $\gamma_c$ and $d$ characterize polariton decay rate and the diameter of the pillar, respectively. The derivation of nonlinearity and noise rates for the polariton model are presented in Appendix~\ref{Sec:AppendixA}.

We subdivide the period of switching process $T_\mathrm{tr}$ into three parts of equal duration. At first, we gradually increase  by $i\sigma$ the on-site potential of two last sites of the boundary unit cell $n_b$, which is the leftmost cell in Fig.~\ref{fig:eigensystem}(b). Second, we perform the switching of on-site potentials in the boundary unit cell $n_b+1$. Finally, in the third step we gradually turn off the supporting potential.
Hence, the majority of on-site potentials in Eq. \eqref{eq:dynamics} are time-independent, except the vicinity of domain wall. Particularly, for the unit cell $n_b+1$, the middle cell in Fig.~\ref{fig:eigensystem}(b), the switching is given by
\begin{equation}
    \epsilon_{n_b+1,A[C]}= \left( \pm g_3 + \frac{\pm (g_1-g_3)}{1+e^{-(t-\tau_{1})/\Delta\tau}}  \right) e^{i\theta} -i\gamma,
\end{equation}
\begin{equation}
    \epsilon_{n_b+1,B[D]}= \left( \pm g_4 + \frac{\pm (g_2-g_4)}{1+e^{-(t-\tau_{1})/\Delta\tau}}  \right) e^{i\theta} -i\gamma,
\end{equation}
where $+$ $[-]$ signs correspond to sites (A,B), [(C,D)], respectively. In addition, for the last two sites of unit cell $n_b$ we have
\begin{equation}
    \label{eq:suppC}
    \epsilon_{n_b,C}= -g_1 e^{i\theta} +  \frac{i\sigma } {1+e^{(t-\tau_{2})/\tau}} \left(1-\frac{1}{1+e^{(t-\tau_{0})/\Delta\tau}} \right)   , 
\end{equation}
\begin{equation}
    \label{eq:suppD}
    \epsilon_{n_b,D}= g_2 e^{i\theta} +  \frac{i\sigma} {1+e^{(t-\tau_{2})/\Delta\tau}} \left(1-\frac{1}{1+e^{(t-\tau_{0})/\Delta\tau}} \right)    .
\end{equation}
Here
$\tau_{2}-\tau_{0}=T_\mathrm{tr}$, $\tau_{1}=\tau_{0}+T_\mathrm{tr}/3$,
and $\Delta\tau$ characterizes the transition rate, which we choose as $\Delta\tau \approx T_\mathrm{tr}/45$. The dependence of imaginary part of the on-site potentials is shown in Fig.~\ref{fig:dynamics}.

In order to quantify the efficiency of the transition we calculate the coherence between the initially existing and the newly emerging domain walls.
As mentioned above,  the domain wall is mainly localised in 5 sites around the boundary between two phases.
Accordingly, we introduce a vector describing this state
\begin{equation}
    |\Psi_{n} (t) \rangle= 
    \begin{pmatrix}
        \psi_{n, B}   \\
        \psi_{n, C}   \\        
        \psi_{n, D}   \\
        \psi_{n+1, A} \\
        \psi_{n+1, B} \\
    \end{pmatrix}
    .
\end{equation}
Then the spatio-temporal first order correlation function between domain walls can be defined as
\begin{equation}
    \label{eq:gonedef}
g^{(1)}_{(n_b,\Delta n)}(\tau) = \frac{\langle \Psi_{n_b}^{*}(\tau_0)\Psi_{n_b+\Delta n}(\tau_0 + \tau)\rangle_{\mathcal{N} }}{\sqrt{\langle |\Psi_{n_b}(\tau_0)|^2\rangle_{\mathcal{N} } \langle |\Psi_{n_b+\Delta n}(\tau_0+\tau)|^2\rangle_{\mathcal{N} } } }  ,
\end{equation}
where $\Delta n=0,1,2,\ldots$ corresponds to a shift of the domain wall by respective number of unit cells, $\tau = t - \tau_0$ is the time interval.
Here the average $\langle \dots \rangle_{\mathcal{N} }$ is performed over a large number $\mathcal{N}$ of stochastic realizations. In all our simulations we use $\mathcal{N} = 1000$.

\subsection{Results}

\begin{figure}[t]
	\centering
	\includegraphics[width=\linewidth]{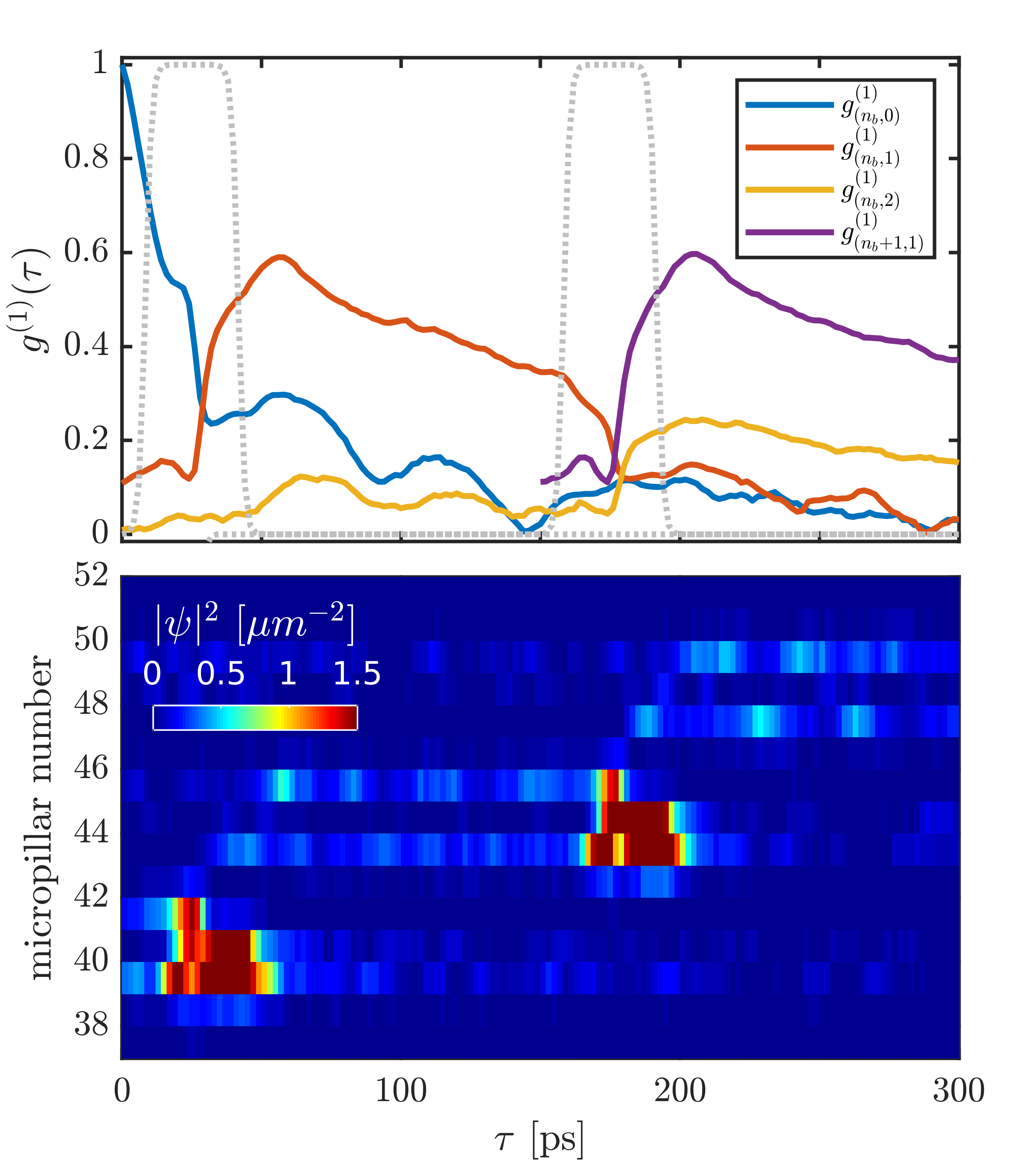}
	\caption{
	The evolution of coherence (top panel) and population (bottom panel) during a two switch transfer. The two switching events are separated by a large temporal window (between 50 and 150 ps), which is necessary for the establishment of steady state after the first switch.
    The blue curve in top panel shows the temporal dependence of spatially local first order correlation function in the initial domain wall. At the end of the saturation window it tends to zero, due to the complete depletion of population in the initial domain wall (see the bottom panel). Some revival during the second switching event is an artefact of stochastic simulations. The red and purple curves correspond to the coherence between initial and intermediate (red), and intermediate and final domains (purple). The very similar shape of these curves indicates that the two switching events are identical. The orange curve shows the correlation function between the initial and final domain walls. 
	The gray dotted lines correspond to temporal profiles of supporting pumps during the switching events.
	}
	\label{fig:doubleswitch}
\end{figure}
\begin{figure}[t]
	\centering
	\includegraphics[width=\linewidth]{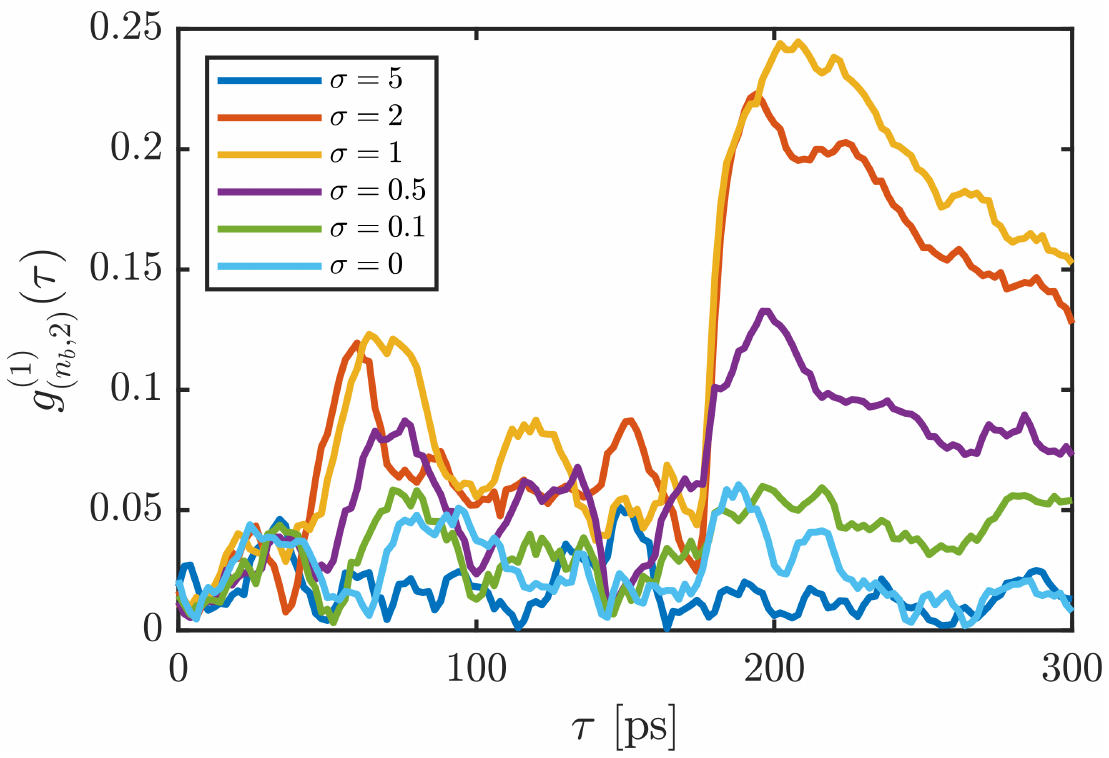}
	\caption{
	Evolution of spatio-temporal correlation function during the shift of domain wall by two cells for different values (in units of $\kappa$) of the amplitude of supporting on-site potential. The enhancement of supporting potential prevents the existing domain wall mode from depletion. On the other hand, it is an additional source of noise. The interplay of these factors determines the optimal rate of supporting potential to be $\sigma=\kappa$.
	}
	\label{fig:amplitude}
\end{figure}

Numerical simulations were performed for the model of an exciton-polariton lattice, where the tunnelling rate is chosen as $\kappa=0.1$ meV. All the other quantities in the Hamiltonian are scaled relative to $\kappa$. Correspondingly, the temporal evolution is presented in ps. The parameter $d$ in the definition of white noise corresponds to the diameter of the micropillar, and is chosen as $d=3$ $\mu$m.
The supporting on-site potential is $\sigma=\kappa$, and the scaling factor of noise amplitude is chosen as $\beta=0.05$, corresponding to characteristic coherence time for polariton condensates \cite{Krizhanovskii2008}.
The dependence of coherence degree on $\beta$ is discussed in Appendix~\ref{Sec:AppendixB}.

In Fig. \ref{fig:dynamics} we present the evolution of the first order correlation function of~\eqref{eq:gonedef} during a shift of a domain wall by one unit cell,  for different switching rates. We start at $t=0$ and perform the evolution for 500 ps with time independent Hamiltonian, during which we reach the nonequilibrium steady state. 
Note that from here onward, in all figures this part of the evolution is not shown. Then we gradually change the potentials in the boundary unit cells $n_b$ and $n_b+1$.
The process is analogous to Landau-Zener transition \cite{Landau1932,Zener1932}. 
One can expect that the lower is the transition speed, the higher will be the preserved degree of coherence due to a smaller perturbation of the steady state. 
This indeed is the case when increasing the switching time $T_\mathrm{tr}$ from 1 to 50 ps. However, due to non-Hermiticity of the Hamiltonian, the system is intrinsically of open-dissipative nature. This circumstance imposes a finite coherence time even in the steady state, stemming from the noise associated with input and output channels. In the context of exciton-polaritons these channels are represented by external incoherent pump and the finite lifetime of polaritons. 
Correspondingly, the finite coherence time imposes the upper limit on the transition period.
Thus, the interplay of these two factors determines the timescale optimal for coherent transfer of the domain wall state to the next unit cell.

We further analyse the impact of the supporting on-site potential applied during the transfer protocol, defined by Eqs. \eqref{eq:suppC}, \eqref{eq:suppD}.
In Fig. \ref{fig:protocols} we present the temporal evolution of the coherence degree in the presence and absence of the potential. First of all we study coherence in the absence of switching, i.e. $g_{n_b,0}(\tau)$. 
Blue and red curves correspond here  to $g_{n_b,0} (\tau)$ with and without applying supporting potential, respectively.
We note that the additional noise stemming from the supporting potential largely reduces coherence.
On the contrary, the domain coherence in the case of shift by one unit cell  is found to be very small in the absence of the supporting potential (see orange curve) whereas its presence allows to essentially preserve the coherence degree (purple line). 
It is remarkable that after the switching event, the degree and temporal evolution of the shifted domain coherence (purple curve) is very close to that of coherence in the absence of switching (blue curve). This is another clear indication of the coherent nature of the domain wall transfer event.

In addition, we consider temporal dependence of the first order correlation function in the case of two consequent switching events, as shown in Fig. \ref{fig:doubleswitch}. 
The two switching events are separated by a large temporal window, during which the initial domain wall becomes depleted, and the newly established domain wall reaches to quasi steady state. Interestingly, the temporal profiles of the correlators $g_{n_b,1}$, $g_{n_b+1,1}$ [red and purple curves in Fig.~\ref{fig:doubleswitch}(a)] are almost identical, indicating that the two switching events have same nature. Finally, after the second transition the correlator $g_{n_b,2}$ [orange curve in Fig.~\ref{fig:doubleswitch}(a)] reaches the maximum value of 0.25. Evidently, this value shows that for the chosen parameters the multiple transitions will completely wash out the initial coherence.

In order to increase the conservation of coherence  during the multiple transitions it is necessary the decrease the strength of the noise.
However, in a realistic polariton model the latter has a limited range of allowed values.
The dependence of coherence on scaling of the noise amplitude is discussed in Appendix~\ref{Sec:AppendixB}. 
A possible alternative could be the variation of the strength of supporting on-site potential.
The dependence of domain coherence on the strength of supporting potential is shown in Fig. \ref{fig:amplitude}.
The enhancement of supporting potential amplitude prevents the existing domain wall mode from depletion,
increasing thus the conservation of coherence.
On the other hand, the stronger the pump, the more noise is introduced into the system, decreasing the coherence. The interplay of these factors determines the optimal rate of supporting potential $\sigma=\kappa$, which is used in majority of the calculations.

\section{Conclusion}

In conclusion, we show the possibility of controllable coherent transport of topological domain walls in a system of coupled Bose-Einstein condensates. We found that topologically protected domain wall emerges solely due to spatial modulation of complex-valued on-site potential, while its temporal modulation causes the transfer of domain wall. As a toy model we employ a system of interacting exciton-polariton condensates, where the on-site potential stems from the incoherent pump, and its spatial modulation provides the topological protection. 
We demonstrate that for the high purity systems with low noise amplitude a substantial coherence rate can be retained within a long-range transfer path, being an important prerequisite for practical applications.

\section*{Acknowledgements}

We acknowledge support from the National Science Centre, Poland grant No.~2016/22/E/ST3/00045 and grant No.~2017/25/Z/ST3/03032 under QuantERA program. VS acknowledges support from the mega-grant No. 14.Y26.31.0015 of the Ministry of Education and Science of the Russian Federation.


\appendix
%


\section{THE POLARITON MODEL}
\label{Sec:AppendixA}

A possible system for the realization of the proposed phenomena is a one-dimensional (1D) lattice of coupled micropillars \cite{Comaron2019}. Each micropillar contains a quantum well and is assumed to host a tightly bound exciton-polariton mode. In the mean-field approximation the evolution of the system can be described by discrete mean-field Gross-Pitaevskii equations
\begin{align}
    \label{eq:GPE}
  i \hbar \dot{\psi}_{n,i}&= -\kappa\sum_{\langle nn \rangle} \psi_{n',i'} + \left[g_R n_{n,i}^R+i \frac{Rn_{n,i}^R-\gamma_c}{2}\right]\psi_{n,i},\nonumber \\
  \dot{n}_{n,i}^R&=P_{n,i} - \left(\gamma_R + R |\psi_{n,i}|^2\right)n_{n,i}^R,
\end{align}
where $\psi_{n,i}(t)$ is the condensate amplitude in the $n$-th lattice cell, $n_{n,i}^R(t)$ is the density of exciton reservoir in the $n,i$-th site, $P_{n,i}$ is the external nonresonant pumping rate, $\gamma_c$ and $\gamma_R$ are the decay rates of the condensate and the reservoir, respectively, $g_c$ and $g_R$ are the corresponding interaction constants, and $R$ is the  rate of scattering from the reservoir to the condensate. We assume that the polariton interactions within the condensate are negligible in comparison with the reservoir-condensate interaction $g_Rn_{n,i}^R$, which is a good approximation in most experiments where nonresonant pumping is used.

\begin{figure}[t]
	\centering
	\includegraphics[width=\linewidth]{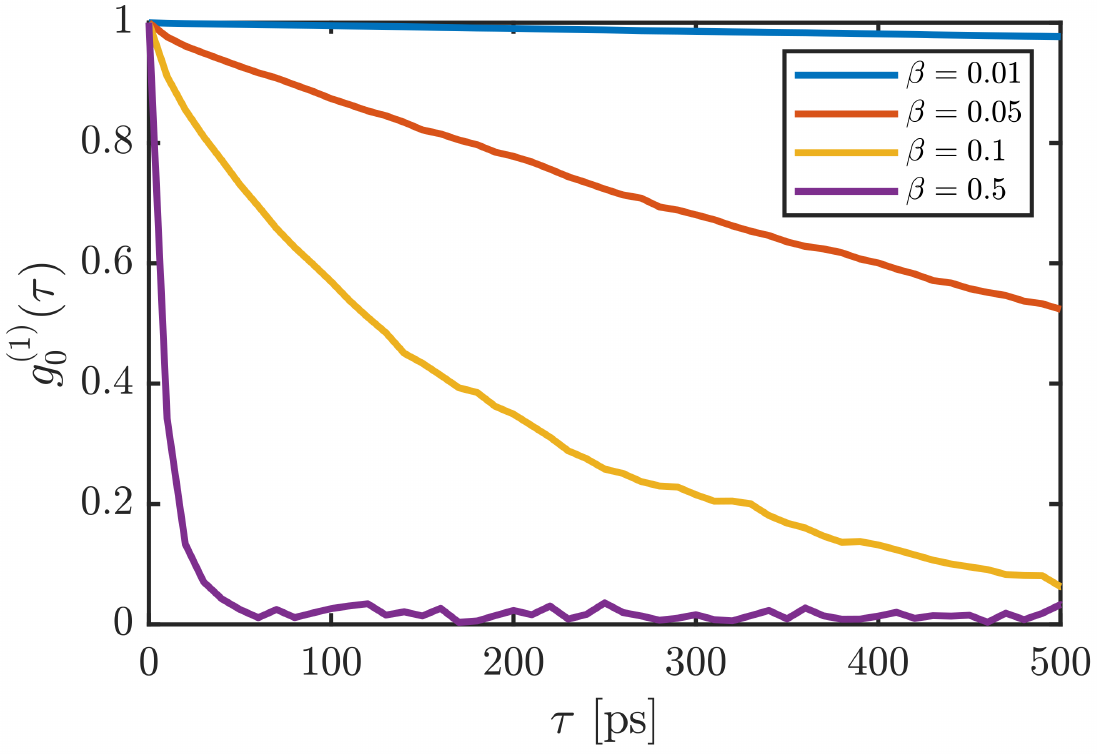}
	\caption{
	The temporal evolution of condensate coherence in a single pillar 
	defined by Eq. \eqref{eq:g0},
	for different scaling of the noise amplitude $\beta$.
	}
	\label{fig:noise}
\end{figure}

In the adiabatic approximation~\cite{Bobrovska2015Adiabatic} we can write
\begin{equation}
    n_{n,i}^R = \frac{P_{n,i}}{\gamma_R+R|\psi_{n,i}|^2}\approx \bar{n}_{n,i}^R  - \frac{R}{\gamma_R} \bar{n}_{n,i}^R |\psi_{n,i}|^2 + O(|\psi_{n,i}|^4),
\end{equation}
where $\bar{n}_{n,i}^R =P_{n,i}/\gamma_R$. 
The on-site potential in Eq. \eqref{eq:dynamics} stems from the linear terms in Eq. \eqref{eq:GPE}. 
In particular, we have
\begin{align} 
    &g_{n,i} e^{i\theta} -\gamma(i+\tan\theta) = g_R \bar{n}_{n,i}^R + \frac{i}{2}\left(R \bar{n}_{n,i}^R - \gamma_c\right) \nonumber\\
    &= \left(\bar{n}_{n,i}^R - \frac{\gamma_c}{R}\right)\left(g_R+\frac{i}{2}R\right) + \frac{g_R\gamma_c}{R}.
    \label{eq:epsilons0}
\end{align}
The last term in the above expression is a constant real energy shift, which can be removed by introducing a rotating frame for condensate amplitudes, $\psi_{n,i}\rightarrow\psi_{n,i}{\rm e}^{-i(g_R\gamma_c/R)t}$. 
Correspondingly, the term $\gamma\tan\theta$ is not present in Eq.~\eqref{eq:dynamics} of the main text, as it only leads to irrelevant uniform energy shift.
Thus, we introduce the following notations for linear and nonlinear terms
\begin{align} 
    &g_{n,i} e^{i\theta} -i\gamma \equiv  \left(\bar{n}_{n,i}^R - \frac{\gamma_c}{R}\right)\left(g_R+\frac{i}{2}R\right) ,
    \label{eq:epsilons}
\end{align}
and
\begin{equation}
    \Gamma_{n,i} \equiv \frac{R}{\gamma_R} \bar{n}_{n,i}^R,
    \label{eq:Gammas}
\end{equation}

In the first of Eqs.~\eqref{eq:GPE} pumping of the condensate is represented by the term $R n_{n,i}^R$, and losses by the decay rate $\gamma_c$.
Considering quantum fluctuations only, the noise density is the sum of noise associated with these channels. Using Eq.~\eqref{eq:epsilons}, one has

\begin{align}
    R \bar{n}_{n,i} + \gamma_c &= 2 \left[ \left(\bar{n}_{n,i}^R - \frac{\gamma_c}{R}\right)\frac{R}{2}+ \gamma_c \right]
    \notag \\
    &    = 2 \left[ g_n \sin\theta -\gamma + \gamma_c \right] ,
\end{align}
Thus, we obtain the density of quantum noise as
\begin{equation}
     \langle \xi_{ni} (t) \xi^{*}_{n'i'} (t') \rangle =  \delta_{nn'}\delta_{ii'} \delta_{tt'}   
    \beta^2 2 \left[ g_n \sin\theta -\gamma + \gamma_c \right] /d .
    \label{eq:noise}
\end{equation}
Here $d$ denotes the diameter of the micropillar, and $\beta$ is dimensionless scaling parameter. The parameter $\beta$ is introduced here because the above simple theoretical argument overestimates the amplitude of quantum fluctuations. In our work, $\beta$ is chosen to match the results of our simulations for the condensate coherence decay time with those reported in experimental investigations~\cite{Krizhanovskii2008}, i.e. the order of several hundreds of ps.

Finally, substituting the Eqs.~\eqref{eq:epsilons}, \eqref{eq:Gammas} into Eq.~\eqref{eq:GPE}, and adding the noise term \eqref{eq:noise}, we reach at the Eq.~ \eqref{eq:dynamics} of the main text.

\section{THE SCALING OF THE NOISE AMPLITUDE}
\label{Sec:AppendixB}

Here we briefly discuss the dependence of first-order correlation function on the noise scaling factor $\beta$. For that, we study the temporal coherence of condensate in a single micropillar, defined as 
\begin{equation}
    g_0^{(1)} (\tau)= \frac{\langle \psi_{n,i}^{*}(\tau_0) \psi_{n,i}(\tau_0 + \tau)\rangle_{\mathcal{N} }}{\sqrt{\langle |\psi_{n,i}(\tau_0)|^2\rangle_{\mathcal{N} } \langle |\psi_{n,i}(\tau_0+\tau)|^2\rangle_{\mathcal{N} } } },
    \label{eq:g0}
\end{equation}
$\forall (n,i)$, and
in the absence of hopping between the pillars, i.e. $\kappa=0$.
Fig.~\ref{fig:noise} shows evolution of coherence for different values of $\beta$. As expected, the coherence decays exponentially, with the decay rate increasing together with noise amplitude.


\bibliography{biblio}

\end{document}